\documentclass{chjaa}                  

\usepackage{graphicx,times,rotating}             
\input{epsf.sty}                        
\input{psfig.sty}                       

\begin{document}

   \title{A MODIFIED STRATIFIED MODEL FOR 3C 273 JET
}

   \volnopage{Vol.0 (200x) No.0, 000--000}      
   \setcounter{page}{1}           

   \author{Wen-Po Liu
      \inst{1,2}
   \and Zhi-Qiang Shen
      \inst{1}
      }

   \institute{Shanghai Astronomical Observatory, Shanghai 200030,
P.R.China; wpliu@shao.ac.cn; zshen@shao.ac.cn\\
        \and
             The Graduate School of Chinese Academy of Sciences, Beijing 100049,
P.R.China\\
          }


   \abstract{
We present a modified stratified jet model to interpret the observed
spectral energy distributions of knots in 3C 273 jet. Based on the
hypothesis of the single index of the particle energy spectrum at
injection and identical emission processes among all the knots, the
observed difference of spectral shape among different 3C273 knots
can be understood as a manifestation of deviation of the equivalent
Doppler factor of stratified emission regions in individual knot
from a characteristic one. The summed spectral energy distribution
of all the ten knots in 3C 273 jet can be well fitted by two
components, low-energy (radio to optical) component dominated by the
synchrotron radiation and high-energy component (UV, X-ray and
$\gamma$-ray) dominated by the inverse Compton scattering of the
cosmic microwave background. This gives a consistent spectral index
of $\alpha=0.88$ ($S_\nu \propto \nu^{-\alpha}$) and a
characteristic Doppler factor of $7.4$. Assuming the average of the
summed spectrum as the characteristic spectrum of each knot in the
3C273 jet, we further get a distribution of Doppler factor. We
discuss the possible implications of these results for the physical
properties in 3C 273 jet. Future GeV observations with $\it GLAST$
could separate the $\gamma$-ray emission of 3C 273 from the large
scale jet and the small scale jet (i.e. the core) through measuring
the GeV spectrum.
   \keywords{galaxies: active --- galaxies: individual (3C273)
--- galaxies: jets --- radiation mechanisms: non-thermal}
   }

   \authorrunning{W.-P. Liu \& Z.-Q. Shen}            
   \titlerunning{A MODIFIED STRATIFIED MODEL FOR 3C 273 JET}  

   \maketitle

%
%
\section{INTRODUCTION}           

Being nearby (z=0.158), the jet of 3C 273 which was the first quasar
discovered (Schmidt 1963) has been studied extensively from the
radio (e.g., Conway et al. 1993), infrared (e.g., Jester et al.
2005; Uchiyama et al. 2006, hereafter U06; Jester et al. 2007,
hereafter J07; Wen et al. 2002), optical (e.g., Jester et al. 2001;
2005; Lin 2006; Qian 2001), to the X-rays (e.g., Marshall et al.
2001; Sambruna et al. 2001, hereafter S01; Jester et al. 2006,
hereafter J06). Chaotic feature in the light curve of 3C 273 is also
discussed by Liu (2006). A TeV flux upper limit has also been
obtained by shallow $High ~Energy ~Stereoscopic ~System$ (H.E.S.S.)
observations (Aharonian et al. 2005). The $10^{\prime\prime}$-long
radio jet has a knotty morphology, with the first bright knot at
about $12^{\prime\prime}$-$13^{\prime\prime}$ from the central
engine and increasing radio intensity toward a terminal bright hot
spot about $22^{\prime\prime}$-$23^{\prime\prime}$ from the nucleus
(Flatters $\&$ Conway 1985; U06).

Based on the spectral energy distributions (SEDs) of these knots and
hot spots in 3C 273 jet, many researchers have identified
two-component nature (U06; J06; S01), namely (1) the low-energy
component extending from radio to optical, and (2) the high-energy
component responsible for the emission including X-ray. Synchrotron
emission is considered to be the dominant radiation mechanism from
radio through optical bands, i.e., the low-energy component (S01;
U06), but the radiation mechanism of X-ray emission is still
perplexing. So far, there have been three candidates for the origin
of the X-rays (S01): (a) synchrotron emission which another much
more energetic population of particles emit (R\"{o}ser et al. 2000;
Aharonian 2002; Bai $\&$ Lee 2003; U06); (b) inverse Compton
scattering of synchrotron photons from low energy component, i.e.
Self-Synchrotron Compton (SSC) model; and (c) inverse Compton
scattering of photons which are external to the jet£¬ such as
inverse Compton scattering of cosmic microwave background photons
model (hereafter IC/CMB).

SSC emission from electrons in an equipartition magnetic field can
usually account for the X-ray emission from hot spots (J07, Harris
$\&$ Krawczynski 2006). But S01 and J06 showed that the contribution
from SSC X-rays in 3C 273 jet is not significant, so we
provisionally ignore SSC effect on the X-ray emission.
Georganopoulos et al. 2006 (hereafter G06) have presented some
diagnostics (the synchrotron model needs multi-TeV electrons
responsible for production of the X-rays, but the IC/CMB model
requires a cutoff which is lower than TeV in the SED) to distinguish
the synchrotron and IC/CMB models. And the $\gamma$-ray predicted by
these two models, can be tested through GeV and TeV observations of
the 3C 273 large-scale jet. U06 described the wide radio-to-X-ray
spectrum with their double power-law models (Eq. (1) of U06,
low-energy component with an exponential cutoff), but they required
an additional assumption that the two power-law indices were
different and their formula was partly man-made rather than derived
from any clear physical insight. The double synchrotron emission
model also does not clearly point out the direct connection of the
two populations of source particles, and the second, the
ultra-energetic particles that emit X-ray photons calls for a
radical rethink of the physics of relativistic jet that black holes
drive (Urry 2006).

Among the suggested mechanisms for the high-energy component, we
adopt IC/CMB model if there is a bulk relativistic motion on
kiloparsec scales (S01). Indeed, VLBI observations have detected
apparent superluminal motions in the parsec-scale jet with apparent
velocities $\sim6$-$10$ c (e.g., Unwin et al. 1985). And
X-ray/optical emission remains asymmetric, implying the presence of
relativistic bulk motion on kiloparsec scales (S01) too. J06
presented deeper Chandra observations of 3C 273 jet and found that
the X-ray spectra are softer than the radio spectra in nearly all
parts of the jet, ruling out the simplest one-zone beamed IC/CMB
models for the X-ray emission from the entire jet. Within their
two-zone jet model, they still required two different spectral
indices for radio and X-ray spectrum of each knot in 3C 273 jet. G06
sum the fluxes of 3C 273 knots excluding the bright knots A and B as
a ``big knot", and they have predicted the $\gamma$-ray emission of
the 3C 273 large-scale jet by applying (not fitting) IC/CMB model to
the ``big knot". S01 divided all the detected knots into four
regions, and then quite well fit IC/CMB model to the wide-band SED
in each of them, implying the rationality of IC/CMB model for 3C273
jet. Then the question is why IC/CMB model could fit the regions
enclosing some neighboring knots but not individual knot. S01, J06
and G06 did not explain this interesting problem. This problem may
be related to the small viewing angles of knots with the stratified
structure (see $\S$ 2). As a result, the observed spectral shape may
be modulated by different beaming effects on the flux. S01's
enclosing some neighboring knots thus minimizes the deviation of
spectrum from that of a single component.

By now, all the fitted models on the SEDs of knots in 3C 273 jet
have assumed a single Doppler factor for emission regions in the
3C273 knots and different spectral indices for the inner and outer
knots (U06; J06; S01). Based on the analysis of Liu $\&$ Shen (2007)
(hereafter LS07) on the radiation mechanism of knots in the M87 jet,
the fitted energy spectral indices of the source particles along the
jet were nearly constant, suggesting the identical acceleration and
radiation mechanism for the source particles along the jet. We would
try a different way (partly similar to S01 and G06's management of
enclosing some neighboring knots) to investigate what resulted in
the observed SEDs of knots in 3C 273 jet.

In $\S$ 2, we describe in detail our model and management for SEDs
of knots in 3C 273 jet. In $\S$ 3, we present and discuss the
fitting results of this model to the summed SED of all the knots in
3C 273 jet and give the distribution of the Doppler factor in the
emission regions among all the knots. A summary is given in $\S$ 4.

\section{A MODIFIED STRATIFIED JET MODEL AND MANAGEMENT FOR THE SPECTRA OF KNOTS}

The effect of the stratified emission along the jet and the effect
of different emitting regions through the jet's cross section are
common for the observed knot emission (Perlman et al. 1999, Marshall
et al. 2002, Perlman $\&$ wilson 2005, Harris $\&$ Krawczynski
2006). Based on the difference between the optical and radio
polarimetric observations, Perlman et al. (1999) advanced a model of
partial energy stratification. They illustrated that the
optical-emitting electrons are located closer to the jet axis, while
most of the radio-emitting electrons are located nearer the jet
surface. Further promoting this model, we think that even the
optical- and radio-emitting electrons themselves may also be
stratified through the jet's cross section, similar to the
spine-sheath structure.

And the advection and diffusion of the populations of particles
along the jet with the decrease of the synchrotron lifetime from low
energy to high energy would result in spatially stratified (LS07)
and conical emission layers (regions) along the jet. Because of
cooling process, the stratified effect along the jet may be more
obvious for high energy electrons. This scenario suggests that the
observed different radiation from a knot at different frequency is
not from a simple electrons component within the same region of that
knot, but from a more complicated one whose stratified emission
layers roughly correspond to different synchrotron-emitting
electrons (and so to different emissions). This is more significant
for a jet with small viewing angle. If the stratified emission
layers were almost independent on each other (i.e., different
emission regions could have different intrinsic velocities and/or
electrons distribution), their Doppler factors may be different.
Otherwise the Doppler factors should be similar to each other. If we
further assume that all the conical emission layers had a same
intrinsic velocity but electron distribution of each emission layer
was asymmetrical to the jet axis, this results in that the
equivalent/averaged viewing angle of each emission layer could have
two phases, i.e., near and away from line of sight.

The model considered here is basically composed of two components
with synchrotron radiation dominating the low-energy band and the
IC/CMB dominating the high-energy emission. We adopted the
synchrotron model of LS07 to describe the low-energy component. In
the model (Eq. (3) of LS07), they considered a decay of spectral
index of injection particles possibly due to the sum of the
injection spectrum from different acceleration sources with
synchrotron losses in the thin acceleration region, so there are two
break frequencies (Eq. (5) in LS07) at two sides of which the
spectral index changes for the spectra of knots in AGN jet. The
IC/CMB spectrum is an identical copy of the synchrotron one. But
since 3C 273 jet in high-energy component was only detected from UV
to X-ray, we use the power law distribution to describe the spectrum
below the first break frequency which falls in $\gamma$-ray band.

Our modified stratified jet model showed that the stratified
emission layers in a knot may have different equivalent/averaged
Doppler factors (i.e., Doppler factors may be roughly a function of
synchrotron frequency). Especially, when the bulk velocity of
advection in 3C 273 jet approaches to the light speed and the
equivalent/averaged viewing angle of emission layers is very small,
both the Doppler factor and apparent motion of emission regions in
the 3C273 knot are sensitive to the change of velocities and
equivalent viewing angles of emission layers. For synchrotron
emission, the observed fluxes ($S_\nu \propto \nu^{-\alpha}$,
$\alpha$ is spectral index) at different frequencies are related to
their Doppler factors by
$S_{\nu_1}/S_{\nu_2}=(\delta_1/\delta_2)^{3+\alpha} \times
(\nu_1/\nu_2)^{-\alpha}$ (Dermer 1995), and for IC/CMB emission, the
relation is $S_{\nu_1}/S_{\nu_2}=(\delta_1/\delta_2)^{4+2\alpha}
\times (\nu_1/\nu_2)^{-\alpha}$ (Dermer 1995). This flux ratio is
very sensitive to the ratio of the Doppler factors, implying that
different bulk velocities and the equivalent viewing angles (i.e.,
different equivalent Doppler factors) can easily cause the
difference in the spectral shape of knots in 3C 273 jet.

Because we do not know the real Doppler factor distribution for
emission regions along 3C 273 jet, in our analysis we assumed that
the deviations of observed spectral shape for emission regions along
3C 273 jet is somehow symmetric to a hypothetic characteristic one
and then we obtain such a characteristic one by averaging SEDs of
all the knots in 3C 273 jet. This is justified by the reasonable
fitting results of the spectral shapes and Doppler factors (see $\S$
3). As such, the increasing lower-energy emissions and decreasing
high-energy emissions with the increasing separation from the
nucleus (U06) can be ascribed to the deviation of Doppler factor of
each knot from the characteristic one in 3C 273 jet. The detailed
distribution of magnetic field and electron velocity directions may
also affect the observed spectral shape of knots, but for
simplicity, we do not consider this case. The SED of individual knot
in the M87 Jet (with a lager viewing angle than 3C273 Jet) could be
fitted by a continuous injection (CI) synchrotron with a break power
law (LS07), but this is not the case in 3C273. By assuming the
effects of intrinsic conditions (such as volume, kinetic age,
electron normalization constant and distribution of magnetic field
and electron velocity directions etc.) of jets in M87 and 3C273 are
similar on the observed synchrotron spectrum of individual knot, we
think that the deviation of the observed spectral shape of
individual knot in 3C 273 jet from the CI synchrotron one is caused
by some external effects likely being related to the different
viewing angles. According to this viewpoint, the spectrum of every
knot in 3C 273 jet is modulated by different equivalent Doppler
factors in the stratified emission layers, resulting in the models
with a single Doppler factor to be ineffective for knots (such as
IC/CMB in J06). But the summed spectrum of all the knots may
minimize such an effect (partly similar to S01' management of
enclosing some neighboring knots) and therefore can be treated as a
representative SED of a ``big knot" with a single Doppler factor and
spectral index. In the following, unlike G06's illustration (not
fitting) of a ``big knot" (excluding the bright knots A and B), we
will first get a quantitative description of the characteristic
spectrum of our ``big knot" (including all the 3C 273 knots) by
means of two-component fitting. Then, we will discuss the deviation
of Doppler factors in each knot by comparing their SEDs with this
characteristic one.

\section{FITTING RESULTS AND DISCUSSION}

In Table 1, the data of knots except knot H1 were from J07. For knot
H1, the $VLA$ and $HST$ data are from Jester et al. (2005), and the
$\it Spitzer$ data from U06. It was not detected by $\it Chandra$ at
X-ray and other three $HST$ bands. The error bars for the radio,
optical and X-ray data were separately estimated by increasing
$5$-$20$ times the r.m.s noise in J07, within the range of the flux
errors ($1\%$-$5\%$ in all cases for both the radio and optical
measurements, less than $10\%$ for most of the X-ray measurements,
U06). A formal fractional uncertainty of $10\%$ is assigned to the
$\it Spitzer$ data to account for systematic errors of $2\%$-$10\%$
(U06). By adding up the flux densities and error bars of all the
knots in 3C 273 jet, we obtained the SED for the
$10^{\prime\prime}$-long jet (as a ``big knot"). We then performed
the weighted least-squares method to fit the two-component model to
this SED.

Before fitting, we need to divide the available 10 data points from
radio to X-rays into two groups to match the corresponding low- and
high-energy components. By examining the shape of the SED in Fig. 1,
it is likely that the cutting frequency between the two components
is around $1.0\times10^{15}$ Hz. This makes it difficult to decide
the dominant mechanism for the $HST$ measurement at
$1.0\times10^{15}$ Hz. Therefore, we first did two fits of the
low-energy synchrotron emission to the data sets with and without
$1.0\times10^{15}$ Hz measurement, respectively. We found that the
fitting with this $\it$$HST$ data point gave a 3-4 times larger
reduced chi-squares ($\chi^2_{\nu}$). Thus, we chose to interpret
the emission at $1.0\times10^{15}$ Hz being dominated by the IC/CMB.
So, the cutoff frequency between the upper limit frequency of the
low-energy component and the lower limit frequency of the
high-energy component is within $4.85\times10^{14}$ to
$1.00\times10^{15}$ Hz.

Once we fixed such a cutoff frequency, we can perform independent
spectral fitting to both low- and high-energy components as
described in $\S$ 2. In practice, we fit the synchrotron model (Eq.
(5) of LS07, with 4 parameters) to the 7 data points at frequency
lower than $1.0\times10^{15}$ Hz. In this process, a power-law form
is chosen near the break frequencies because we only need to be
concerned with the trend of the break frequencies. There are two
break frequencies in our model, but we do not know exactly where
they are when we fit to the chosen data in Fig. 1. So we first
arbitrarily divide observational data into three groups to perform
the fitting and calculate the corresponding $\chi^2_{\nu}$ by
changing the division. All the possible combinations are tried
before we obtain the best fit with a minimal $\chi^2_{\nu}$ among
them. For the high-energy component fit to the remaining 3 high
frequency data points, only a power law (with 2 parameters) is
needed. In Table 2 are shown the fitting results. The SED for the
``big knot" in 3C 273 jet and the best fits are plotted in Fig. 1.
Therefore, both the high and low-energy components are independently
described by our model. The fitted spectral indices of the
low-energy ($0.87\pm0.02$) and high-energy ($0.88\pm0.02$)
components are well consistent with each other, supporting the
IC/CMB nature of the high-energy component. The corresponding value
of the particle spectral index ($p=2\alpha+1$) of 2.76 is also among
the characteristic range ($2<p<3.5$, Blumenthal $\&$ Gould 1970).

\begin{sidewaystable} \caption[]{Flux Densities of Knots in 3C 273 jet.
\footnotemark[1]{The data of the knots except knot H1 were from J07.
The $\it VLA$ and $\it HST$ data of knot H1 were from Jester et al.
(2005), but the $\it Spitzer$ data of knot H1 were from U06.}
\footnotemark[2]{The error bars for the $\it VLA$, $\it HST$ (except
for $1.86\times10^{15}$ Hz) and $\it Chandra$ data were separately
estimated by increasing $5$-$20$ times of r.m.s noise in these data
from J07.} \footnotemark[3]{A formal fractional uncertainty of
$10\%$ is assigned to the $\it Spitzer$ data to account for the
systematic errors of $2\%$-$10\%$ (U06).} \footnotemark[4]{The sum
of the flux densities and error bars for the knots in the 3C273
jet.}} \scalebox{0.8}{
\begin{tabular}{crrrrrrrrrrr}

\hline\hline
\\
\multicolumn{1}{c}{} & \multicolumn{10}{c}{Flux Density\footnotemark[1]}\\
\cline{2-11}\\
Frequency (Hz) & A & B1 & B2 & B3 & C1 & C2 & D1 & D2H3
& H2 & H1 & Sum\footnotemark[4]\\
\\
\hline
\\
$\it VLA$\footnotemark[2] ($mJy$):\\
$8.33\times10^{9}$
& $90.5\pm1.30$ & $69.2\pm1.07$ & $105\pm1.43$ & $50.8\pm1.01$ & $101\pm1.56$ & $205\pm2.47$ & $283\pm3.9$ & $836\pm8.84$ & $1330\pm16.90$ & $571.9\pm7.27$ & $3400.5\pm45.75$\\
$1.5\times10^{10}$
& $58.9\pm0.66$& $41.9\pm0.53$ & $69\pm0.77$ & $34.8\pm0.59$ & $67\pm0.90$ & $134\pm1.56$ & $182\pm2.34$ & $516\pm5.33$ & $782\pm9.62$ & $398.4\pm4.90$ & $2085.6\pm27.20$\\
$2.25\times10^{10}$
& $38.5\pm0.49$ & $30.6\pm0.42$ & $50.8\pm0.60$ & $23.6\pm0.43$ & $48.8\pm0.69$ & $97.1\pm1.12$ & $131\pm1.69$ & $357\pm3.64$ & $520\pm6.24$ & $208.0\pm2.50$ & $1427.4\pm17.82$\\
\\
$\it Sptizer$ ($\mu Jy$):\\
$5.23\times10^{13}$
& $45\pm10$ & $35\pm12$ & $36.2\pm8.1$ & $12.8\pm2.9$ & $98\pm11$ & $89\pm12$ & $154\pm15.4$\footnotemark[3] & $161\pm16.1$\footnotemark[3] & $87\pm12$ & $28\pm11$ & $746\pm110.5$\\
$8.45\times10^{13}$
& $27\pm2.7$\footnotemark[3]& $15\pm2.6$ & $28.8\pm2.88$\footnotemark[3] & $10.2\pm1.02$\footnotemark[3] &36$\pm3.6$\footnotemark[3] & $46\pm4.6$\footnotemark[3] & $80\pm8$\footnotemark[3] & $140\pm14$\footnotemark[3] & $41\pm4.1$\footnotemark[3] & $10\pm2.6$ & $434\pm46.1$\\
\\
$\it HST$\footnotemark[2] ($\mu Jy$):\\
$1.87\times10^{14}$
& $11.3\pm0.48$ & $4.9\pm0.17$ & $10.2\pm0.20$ & $4.84\pm0.12$ & $10.8\pm0.17$ & $18.1\pm0.20$ & $21.7\pm0.19$ & $40\pm0.28$ & $6.97\pm0.19$ & $2.04\pm0.06$ & $130.31\pm2.06$\\
$4.85\times10^{14}$
& $4.93\pm0.12$ & $1.83\pm0.082$ & $3.86\pm0.098$ & $1.37\pm0.059$ & $2.93\pm0.079$ & $3.93\pm0.09$ & $3.56\pm0.08$ & $7.68\pm0.1$ & $1.29\pm0.09$ & $\cdots$ & $31.38\pm0.72$\\
$1.00\times10^{15}$
& $3.2\pm0.17$ & $0.973\pm0.06$ & $1.88\pm0.07$ & $0.507\pm0.04$ & $1.16\pm0.06$ & $1.28\pm0.06$ & $1.09\pm0.05$ & $2.54\pm0.07$ & $0.459\pm0.07$ & $\cdots$ & $13.09\pm0.65$\\
$1.86\times10^{15}$
& $2.03\pm0.11$ & $0.626\pm0.037$ & $1.47\pm0.079$ & $0.439\pm0.028$ & $0.657\pm0.039$ & $0.723\pm0.042$ & $0.679\pm0.04$ & $1.39\pm0.075$ & $0.288\pm0.02$ & $\cdots$ & $8.302\pm0.47$\\
\\
$\it Chandra$\footnotemark[2] ($nJy$):\\
$2.42\times10^{17}$
& $46.5\pm2.70$ & $10.9\pm1.25$ & $20\pm1.65$ & $3.41\pm0.70$ & $4.85\pm0.80$ & $6.25\pm0.90$ & $5.16\pm0.85$ & $7.82\pm1.00$ & $1.3\pm0.45$ & $\cdots$ & $106.19\pm10.30$\\
\hline
\end{tabular}}
\end{sidewaystable}

\begin{figure}
   \centering
   \includegraphics[scale=1.4]{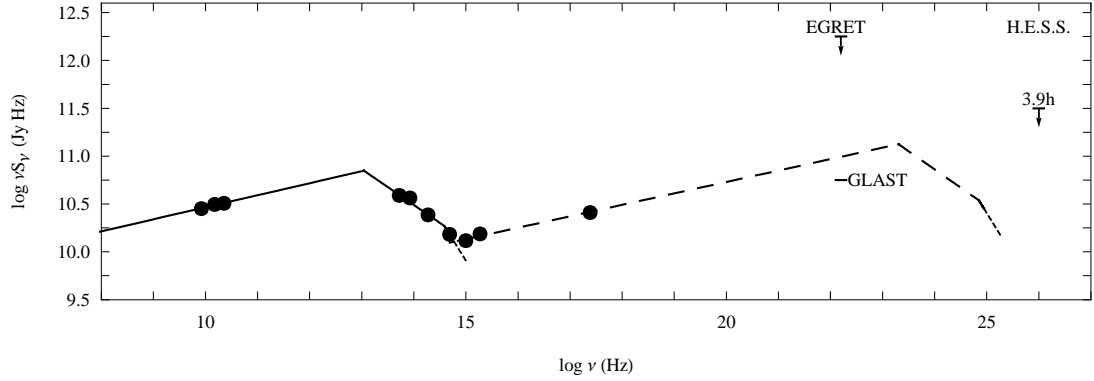}
   \caption{The SED for the summed knots in 3C 273 jet from radio
through X-ray wavelengths. The solid line displays model fit to the
low-energy component, and the dashed line to the high-energy
component. The dotted lines show the possible cutoff frequency band
for the two components. The error bars of most measurements are too
small to be seen here. The EGRET upper limit and the $GLAST$
sensitivity limit are also shown, as well as a TeV flux upper limit
from shallow H.E.S.S. observations (Aharonian et al. 2005).}
   \end{figure}

   \begin{table}
\caption{Parameters of Model Fits to Radio through X-ray Data. Col.
(1): Component designation. Col. (2): Model designation. Col. (3):
Spectral index. Col. (4): Reduced chi square. Col. (5): Peak
frequency, note that the peak frequency for high-energy component is
a derived parameter. Col. (6): Characteristic Doppler factor.}
\begin{center} \scalebox{0.85}{\begin{tabular}{crrrrrrrrrr} \hline
\hline Parameters &  Model & $\alpha$ & $\chi^2_{\nu}$ & $\nu_{peak}$(Hz) & $\delta$ \\
\hline
  Low-energy Component & Synchrotron & $0.87\pm0.02$ & $1.18$ & $1.10\times10^{13}$ & $7.4\pm0.5$\\
  High-energy Component & IC/CMB & $0.88\pm0.02$ & $1.51$ & $2.00\times10^{23}$ & $\cdots$\\
\hline
\end{tabular}}\end{center}
\end{table}

The fitted synchrotron component has a total power of $L_s\approx
5.2\times 10^{41}$ erg s$^{-1}$, which is actually dominated by the
emission around the peak frequency, i.e., $L_s\approx 4\pi D_L^2
[\nu_{s,p}S(\nu_{s,p})/(1-\alpha)+\nu_{s,p}S(\nu_{s,p})/(\alpha-0.5)]$,
here $S(\nu_{s,p})$ is the flux density at the peak frequency
$\nu_{s,p}\approx 1.10\times 10^{13}$ Hz (see Table 2), and a
luminosity distance of $D_L=749$ Mpc is adopted. Accordingly, we can
estimate the luminosity of IC/CMB emission, which is also dominated
by the emission around its peak frequency $\nu_{c,p}$. Thus, we can
get a ratio between IC/CMB luminosity and synchrotron luminosity as
follows
\begin{equation}
{L_c \over L_s}= {\nu_{c,p} \; S(\nu_{c,p}) \over \nu_{s,p} \;
S(\nu_{s,p})}.
\end{equation}
According to Geoganopoulos et al. (2006), this luminosity ratio is
related to the Doppler factor as
\begin{equation}
{L_c \over L_s}= {6.25\times 10^{-4} \; \delta^4},
\end{equation}
and the two peak frequencies ($\nu_{c,p}$ and $\nu_{s,p}$) are
related by the expression
\begin{equation}
{\nu_{c,p}}=3.3 \times 10^8 \; \delta^2 \; \nu_{s,p}.
\end{equation}
It should be mentioned that in getting Eqs. (2) and (3), an
equipartition magnetic field $B\delta\approx 2\times10^{-4}$G
(Jester et al. 2005) is used. These equations require that the
electrons which emit synchrotron photons (i.e., the first component
of the SED) upscatter the CMB and, the resulting IC/CMB SED (i.e.,
the second component) is an exact copy of the synchrotron one (G06).
This is consistent with our model for the large scale jet as a ``big
knot" (our synchrotron model is somewhat different from the one of
G06, but this doesn't influence the validity of the equations in the
letter), so these equations are still valid in our model. Finally,
from our fitting results, we can obtain the characteristic Doppler
factor $\delta\approx7.4$ which well agrees with the requirement of
FSRQ's Doppler factor ($>6.45$, Cao \& Bai 2008). This may verify
the Uniform Scheme of quasars (Urry \& Padovani 1995). The frequency
shift of IC/CMB and synchrotron emissions
($\nu_c\sim1.81\times10^{10}\nu_{s}$), here $\nu_c$ and $\nu_s$ are
the observed IC/CMB and synchrotron frequencies, respectively. Then
we could estimate the equipartition magnetic field $B=27\mu$G and a
synchrotron lifetime of $\sim4000$yr, (Eq. (4) of U06) of the ``big
knot" which is smaller than the source kinetic age
($10^5\sim10^7$yr, U06), this means a continuous injection of
electrons responsible for the SED of the 3C273 jet. The peak
frequency of the high-energy component is at $2.00\times10^{23}$ Hz
$\sim$1 GeV (see Table 2), around the working band of the $Energetic
~Gamma$-$Ray ~Experiment ~Telescope$ (EGRET) on the $Compton
~Gamma$-$Ray ~Observatory$ (CGRO) and the current $Gamma$-$Ray
~Large ~Area$ $\it~Space$ $\it~Telescope$ ($GLAST$). The model
predicted $\gamma$-ray flux was at least 10 times lower than the
EGRET upper limit (Sreekumar et al. 1994; Geoganopoulos et al.
2006), but should be detectable by the $GLAST$ (see Fig. 1).
Although the $GLAST$ (e.g., the expected 68\% containment angular
resolution of $GLAST$ LAT is $0.15^\circ$ (on-axis) for photons
above 10 GeV, $GLAST$ website) could not separate the $\gamma$-ray
emission from the large scale jet and the core, by measuring the SED
of the GeV energies, we still could identify the origin of the GeV
emission (the large scale jet or the small scale jet, i.e. the
core). The high-frequency cutoff of the second component appears at
the band between $8.8\times10^{24}$ and $1.8\times10^{25}$ Hz (lower
than TeV band), which is consistent with the non-detection by
shallow $H.E.S.S.$ observations (Aharonian et al. 2005) as shown in
Fig. 1.

Based on the model described in $\S$ 2., the observed spectrum of
individual knot in 3C 273 jet may result from the stratified
emission layers in which the Doppler factors deviate from the
characteristic one. So, unless removing the deviation effect of
Doppler factors in the observed spectrum of each knot, those fitted
spectral indices for radio and X-ray bands of each knot (Jester et
al. 2005; J06; J07) didn't really reflect the intrinsic and
characteristic spectral shape of 3C273 jet. Here, for comparison, we
take the average of the summed spectrum as the characteristic
spectrum of each knot in 3C 273 jet. Then, through the ratio of the
observed to average flux densities of each knot and the
aforementioned formula in $\S$ 2, we could derive a distribution of
the Doppler factor in the spatially stratified emission regions of
the 3C273 knots (Fig. 2). The error bars of derived Doppler factors
could be evaluated from the ones of observed flux and the
characteristic Doppler factor, and are also plotted in Fig. 2.
Furthermore, by assuming a constant bulk velocity of the emission
layers along the jet (we chose the bulk velocity that satisfies the
average apparent velocity 8c and the characteristic Doppler factor
7.4), we can obtain their equivalent viewing angle distribution of
$5^\circ$ to $11^\circ$ with an average of $7.7^\circ$ from their
Doppler factor distribution. The averaged viewing angle distribution
of emission layers in each knot could be explained by the
combination of different equivalent viewing angle phases (i.e., near
and away from viewing line) of different emission layers in each
knot. As expected, the trend of the Doppler factor distribution is
almost opposite to the equivalent viewing angle distribution.

The derived Doppler factors at radio frequency ($\sim10^{10}$ Hz) in
each knot are almost the same (Fig. 2). This indicated that these
radio-emitting electron regions are not independent. The Doppler
factors at optical emission around $10^{14}$ Hz in 3C273 knots are
distinctly different from each other (Fig. 2), implying that these
optical-emitting electron layers are likely to be independent. As
these optical data are up to the break frequency of 3C273 knots, and
the cooling effect of optical-emitting electrons is more severe than
radio-emitting electrons, then the stratified effect of these
optical-emitting electrons are also more obvious than radio-emitting
electrons.

\begin{figure}
   \centering
   \includegraphics[scale=1.4]{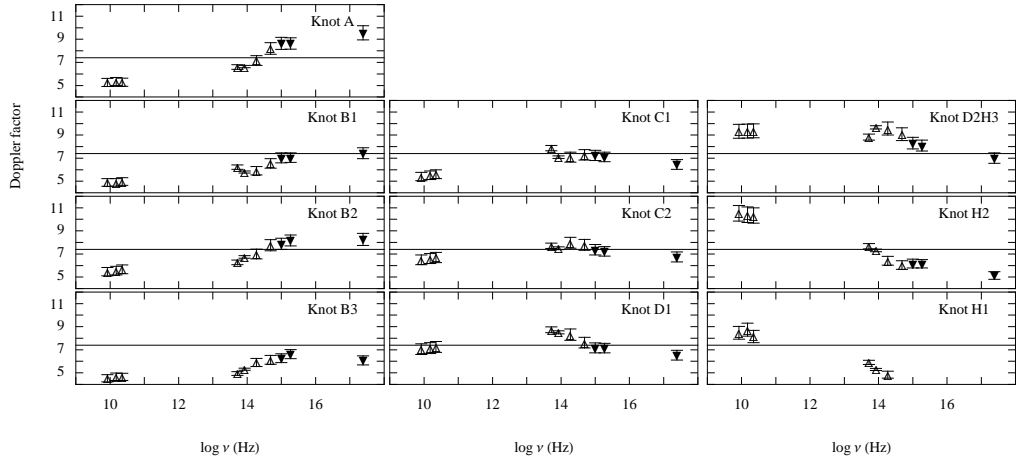}
   \caption{Distribution of the Doppler factor of each knot as a
function of the observing frequency. The low-frequency component
from radio to optical is plotted as empty-up-triangles, the
high-frequency component as filled-down-triangles. The error bars
are also plotted in the data. The left, middle and right panels show
inner, intermediate and outer knots in 3C 273 jet respectively, with
a characteristic Doppler factor of 7.4 in 3C 273 jet indicated by a
straight line.}
   \end{figure}

Based on our fits, the observed spectral shape of X-ray spectrum
($\sim10^{18}$ Hz) in the high-energy component actually reflects
the spectrum of particles that emit photons at a frequency below
$5.51\times10^7$ Hz. Although the low-frequency emissions may be
absorbed severely, the particles in the emission regions still could
scatter off the CMB photons to the X-ray band. So, the Doppler
factors of high-energy component actually represent the ones of
low-energy component. Based on our modified stratified jet model as
described in $\S$ 2, these below $5.51\times10^7$ Hz radio-emitting
electron regions mainly located nearer the jet surface, and are
almost independent of those $\sim10^{10}$Hz radio-emitting electron
regions. Therefore, the Doppler factors of these two bands are also
different. For the inner knots (such as knots A, B1, B2 and B3 in
Fig. 2), the equivalent Doppler factors of the emission regions
where the particles scatter the CMB photons to the X-ray band are
larger than the ones of the emission regions that mainly emit
photons between $10^{10}$ and $10^{15}$ Hz (see Fig. 2). This
explains that the observed SEDs of the inner knots are dominated by
the high-energy component (see fig. 2 of J07 and fig. 5 of U06). For
the outer knots (such as knots D2H3, H2 and H1 in Fig. 2), the
Doppler factors of the emission regions where the particles scatter
the CMB photons to the X-ray band are overall smaller than the ones
of emission regions that mainly emit photons between $10^{10}$ and
$10^{15}$ Hz. This results in the observed SEDs of the outer knots
to be dominated by the low-energy component. The intermediate knots
(such as knots C1, C2 and D1 in Fig. 2) show almost identical
Doppler factors among all the observed emission regions.

\section{CONCLUSION}

Based on the hypothesis of the same intrinsic SED for all the knots
in 3C 273 jet, the difference in their spectral shape is interpreted
as a result of the difference of the equivalent Doppler factors of
our modified stratified emission layers. Further, if we assume a
constant bulk velocity of the emission regions along the jet, the
Doppler factor distribution could be due to the equivalent/averaged
viewing angle distribution for the spatially stratified emission
layers of the knots in 3C 273 jet. Our fitting to the summed
spectrum supports that high energy X-ray emission is dominated by
inverse Compton scattering of the cosmic microwave background. The
predicted $\gamma$-ray spectrum of the large-scale jet in 3C 273
could be further tested by measuring its GeV spectrum from $\it
GLAST$ observations. It should be noted that our model needs the
stratified synchrtron-emitting electrons, which should be verified
by further multi-bands observations.

We thank S. Jester for helpful communications and supplying his data
to us for reference. Many thanks are due to the referee for
constructive comments.

This work has been partially supported by the National Natural
Science Foundation of China (grants 10573029, 10625314, 10633010 and
10821302) and the Knowledge Innovation Program of the Chinese
Academy of Sciences (Grant No. KJCX2-YW-T03), and sponsored by the
Program of Shanghai Subject Chief Scientist (06XD14024) and the
National Key Basic Research Development Program of China (No.
2007CB815405).

\end{document}